\documentclass[twocolumn,preprintnumbers,superscriptaddress,nofootinbib,aps,prd,floatfix]{revtex4}

\usepackage[toc,page,header]{appendix}

\usepackage{titletoc}

\usepackage{epsf,epsfig,subfigure,graphicx,amsmath,amssymb,cancel}
\usepackage{subfigure}
\usepackage{soul}
\usepackage[colorlinks=true,urlcolor=purple,anchorcolor=blue,citecolor=darkgreen,filecolor=blue,linkcolor=red,menucolor=blue]{hyperref}
\usepackage[usenames,dvipsnames]{xcolor}
\usepackage{slashed}
\usepackage{bm}
\usepackage{enumitem}

\usepackage{afterpage}
\usepackage[utf8]{inputenc}
\usepackage{multirow}
\usepackage{makecell}

\allowdisplaybreaks

\def\tr{\mathop{\rm tr}}

\definecolor{DeepPurple}{rgb}{0.294, 0.180, 0.514}
\definecolor{HypotheticBlue}{rgb}{0.216, 0.494, 0.722}
\definecolor{HypotheticGreen}{rgb}{0.302, 0.686, 0.290}
\definecolor{HypotheticOrange}{rgb}{1.000, 0.498, 0.000}

\definecolor{darkgreen}{rgb}{0,0.6,0}
\definecolor{darkorange}{rgb}{0.99,0.5,0}

\usepackage{cleveref}
\crefname{table}{Table}{Tables}
\crefname{equation}{Eq.}{Eqs.}
\crefname{appendix}{App.}{Apps.}
\crefname{section}{Sec.}{Secs.}
\crefname{figure}{Fig.}{Figs.}

\usepackage{tikz}
\usetikzlibrary{decorations.pathreplacing}

\usepackage{ytableau}

\usepackage[compat=1.0.0]{tikz-feynman}

\begin{document}

\vspace{-1cm}

\title{Axiverse Baryogenesis}

\vskip 1.0cm
\author{Pouya Asadi}
\thanks{{\scriptsize Email}: \href{mailto:pasadi@ucsc.edu}{pasadi@ucsc.edu}}
\affiliation{Department of Physics, University of California Santa Cruz and Santa Cruz Institute for Particle
Physics, \\ 1156 High St., Santa Cruz, CA 95064, U.S.A.}
\author{David Cyncynates}
\thanks{{\scriptsize Email}: \href{mailto:davidcyn@ictp.it}{davidcyn@ictp.it}}
\affiliation{Department of Physics, University of Washington, Seattle, WA 98195, U.S.A.}
\affiliation{International Centre for Theoretical Physics, Strada Costiera 11, 34151 Trieste, Italy}
\affiliation{Istituto Nazionale di Fisica Nucleare (INFN), Sezione di Trieste, I-34127 Trieste, Italy}
\author{Stefania Gori}
\thanks{{\scriptsize Email}: \href{mailto:sgori@ucsc.edu}{sgori@ucsc.edu}}
\affiliation{Department of Physics, University of California Santa Cruz and Santa Cruz Institute for Particle
Physics, \\ 1156 High St., Santa Cruz, CA 95064, U.S.A.}

\begin{abstract}
The QCD axion may offer a unified origin for the baryon asymmetry and dark matter through 
axiogenesis. 
However, in the minimal QCD axion scenario, axiogenesis either underproduces baryons or 
overproduces dark matter, and the required kinetic misalignment initial conditions are 
in tension with axion quality. In this \textit{Letter}, we demonstrate that the axiverse 
naturally resolves these tensions: the QCD axion emerges as a linear combination of
multiple axion-like fields, evading the overclosure problem thanks to new dissipation channels, while introducing additional 
Peccei--Quinn symmetries that ensure a high quality QCD axion. 
We illustrate these points in a toy model with two axions. 
This framework predicts a rich phenomenology within experimental reach, including dark matter detection prospects, astrophysical signals, and collider signatures.

\end{abstract}

\maketitle

\vskip 1cm


The near proximity of the baryonic and dark matter (DM) energy densities~\cite{Planck:2018vyg} may suggest a common origin. Given the exponential sensitivity of the baryon asymmetry to ultraviolet (UV) parameters and the strong model dependence of the DM relic abundance, their numerical similarity appears highly non-generic and calls for explanation. This coincidence should thus serve as a guiding principle in constructing viable DM frameworks.

A natural starting point is a \emph{cogenesis} mechanism, in which both the baryon asymmetry and the DM abundance arise dynamically.\footnote{A complete resolution of the coincidence problem requires showing that the framework \textit{generically} yields an $\mathcal{O}(1)$ ratio between visible and dark energy densities across its parameter space.} Embedding a DM candidate within such a unified mechanism not only ties the two relic densities together, but also substantially bolsters the plausibility of the said DM candidate. Motivated by this, we examine whether the QCD axion can participate in cogenesis. The axion was originally introduced to solve the strong CP problem through Peccei--Quinn (PQ) symmetry breaking~\cite{Peccei:1977hh,Peccei:1977ur} and was later recognized as a compelling DM candidate~\cite{Preskill:1982cy,Dine:1982ah,Abbott:1982af}. Moreover, axions were quickly understood to arise ubiquitously in a wide range of UV constructions~\cite{Kim:1979if,Shifman:1979if,Zhitnitsky:1980tq,Dine:1981rt,Witten:1984dg}, making them a natural arena in which to explore cogenesis.

A rotating axion field in the early universe can simultaneously account for the baryon and 
dark matter abundances through the process known as \emph{axiogenesis}~\cite{Co:2019wyp}. 
In this framework, the axion field $\theta \equiv a/f_a$ acquires a large time derivative~\cite{Co:2019jts,Gouttenoire:2021jhk,Eroncel:2024rpe} 
$\dot{\theta}$ which,  through the QCD chiral anomaly and the electroweak 
$B+L$ anomaly, modifies sphaleron \cite{Klinkhamer:1984di} energetics and sources a baryon 
asymmetry \cite{Kuzmin:1992up,Servant:2014bla,Domcke:2020kcp} in the Boltzmann equations for Standard Model (SM) fermions.

However, the original QCD axiogenesis scenario faces two challenges: generating the observed baryon asymmetry requires a large $\dot{\theta}$ at the electroweak  phase transition, but this same motion overproduces axion dark matter by two orders of magnitude~\cite{Co:2019wyp}.\footnote{Astrophobic axion models with suppressed couplings to Standard Model particles have also been investigated as an alternative route to minimal axiogenesis~\cite{Badziak:2023fsc,Badziak:2024szg}.} Moreover, the initial velocity necessary for kinetic misalignment demands explicit PQ breaking, which can jeopardize the axion quality \cite{Kamionkowski:1992mf}. We discuss these tensions in more detail in the companion paper Ref.~\cite{Asadi:2026zbc}.

In this \textit{Letter}, we show that these tensions are naturally avoided in multi-axion models such as the general (\emph{open}~\cite{Petrossian-Byrne:2025mto}) \emph{string} \textit{axiverse}~\cite{Witten:1984dg,Arvanitaki:2009fg}, where numerous axion-like fields, and often additional U(1) symmetries, arise automatically in the UV. In this setting, the QCD axion is a linear combination of many fields, only one of which must exhibit the high PQ quality required for the strong CP solution. The remaining orthogonal directions can naturally experience sizeable PQ-violating effects. Truncating to two axions captures the essential new dynamics enabled by such multi-axion theories while keeping the discussion analytically transparent; see, e{.}g{.}, Refs.~\cite{Cyncynates:2021xzw,Cyncynates:2022wlq,Cyncynates:2023esj,Higaki:2014qua,Kitajima:2014xla,Daido:2015cba,Daido:2015bva,Daido:2016tsj,Ho:2018qur,Nakagawa:2020eeg,Chen:2021wcf,Foster:2022ajl,Murai:2023xjn,Li:2023xkn,Li:2023uvt,Lee:2024toz,Murai:2024nsp,Li:2024okl,Li:2024psa,Li:2025cep,Dunsky:2025sgz,Dessert:2025yvk} for related axiverse constructions.

In this framework, axions that get their mass from \textit{dark confining sectors with heavy vector-like fermions} open new dissipation channels that prevent axion overclosure while preserving successful baryogenesis. The friction sourced by such a sector is naturally small near the electroweak scale, enabling one axion to rotate rapidly during the electroweak phase transition and generate the observed baryon asymmetry. At lower temperatures, however, the friction grows to efficiently drain the excess kinetic energy, thereby resolving the overclosure problem. The additional PQ symmetries also ameliorate the axion quality problem, allowing the explicit PQ violation required for kinetic misalignment to remain consistent with strong CP constraints. 
In this paper, we outline the main features of this construction and its observational implications, in particular its dynamics at and below the electroweak phase transition, while a detailed analysis of the favorable dynamics, both above and below the electroweak scale, are presented in a companion paper~\cite{Asadi:2026zbc}.

\vspace{0.1in}

\paragraph*{\textbf{Sphaleron Transitions ---}}
The connection between axions and fermion asymmetries is rooted in sphaleron transitions \cite{Klinkhamer:1984di}, i.e. topologically non-trivial field configurations that interpolate between gauge vacua distinguished by their Chern–Simons (CS) number, see also Refs.~\cite{Kuzmin:1985mm,Bochkarev:1987wf,Kuzmin:1992up,Rubakov:1996vz,Riotto:1998bt,Servant:2014bla}. 
In this section, we study sphaleron physics and argue that both the generation of fermion asymmetries and the friction exerted on an axion can be understood from their dynamics.

The time derivative of $\theta$ acts as an effective chemical potential for the CS number, e.g., see Refs.~\cite{Servant:2014bla,Dasgupta:2018eha,Domcke:2020kcp}. 
This introduces an effective dependence of the free energy of the thermal bath on the CS vacuum number. 
Likewise, a nonzero fermion asymmetry contributes to the free energy of the thermal bath as well: through the chiral anomaly, transitions between CS vacua generate asymmetries in all Weyl fermions charged under the non-abelian gauge group. 
This asymmetry backreacts on the sphaleron transitions as a wash-out term. 
Consequently, the change in free energy during a sphaleron process is proportional to both $\dot{\theta}$ and the total fermion asymmetries. Thus, the net rate for the sphaleron process at finite temperature can be written as: 
\begin{equation}
 \frac{\alpha_c}{8\pi}\langle\tilde G G\rangle_T =  \dot n_{\mathrm{CS}} \;\propto\; A \, \dot{\theta} \;+\; B \sum_i n_i\propto\dot{\theta} \, ,
    \label{eq:sph_rate}
\end{equation}
where $\langle \cdot \rangle_T$ denotes a thermal average, $G$ is the non-abelian field strength, $\tilde{G}^{\mu\nu} = \frac{1}{2}\epsilon^{\mu\nu\gamma\beta} G_{\gamma\beta} $ its dual, $\alpha_c$ is the gauge structure constant, and $n_{\mathrm{CS}}$ is the CS number density. $n_i$ is the asymmetric abundance of fermion species $i$, summed over all fermions charged under the gauge group, and $A$ and $B$ are proportionality constants. 
The second proportionality of Eq.~\eqref{eq:sph_rate} holds since $\dot{\theta}$ is the only source term in the fermion Boltzmann equations required for fermion asymmetry generation \cite{Sakharov:1967dj}. 
For a detailed review of basic thermodynamics arguments and derivation, see Ref.~\cite{Asadi:2026zbc}.

\begin{figure}
    \centering
    \resizebox{\columnwidth}{!}{
    \includegraphics[width=0.4\linewidth]{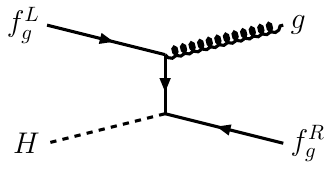}
    \hspace{0.2in}
    \includegraphics[width=0.4\linewidth]{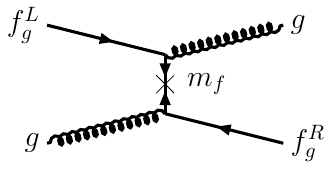}
    }
    \caption{Chirality-flipping process through a Yukawa interaction with a Higgs ($H$) in the thermal bath of the massless gauge boson $g$ (\textbf{left}) or via the mass of the fermion, $m_f$ (\textbf{right}). $f_L$ ($f_R$) denotes the left- (right-)handed fermion. These processes appear as wash-out terms in Boltzmann equations governing fermion asymmetries.}
    \label{fig:chirality}
\end{figure}

Chirality--flipping processes play a crucial role in determining the final fermion asymmetry as well. 
In the electroweak--symmetric phase of the SM, the dominant chirality--flipping interactions arise from scatterings with the Higgs boson (left panel of \cref{fig:chirality}). 
After electroweak symmetry breaking, the leading chirality--flipping process  proceeds via fermion mass insertions (right panel of \cref{fig:chirality}). 
These processes enter the Boltzmann equation for the fermion asymmetries as wash-out terms.

In Ref.~\cite{Asadi:2026zbc} we provide a detailed derivation of these effects and solve the full system of Boltzmann equations to find the final comoving baryon asymmetry~\cite{Co:2020xlh,Domcke:2020kcp}
\begin{equation}
    Y_B \;\simeq\; \frac{45}{2\pi^2 g_{\star S}} \,\frac{3}{158} \,
    \frac{\sum_g \left( 7 y_{u_g}^{-2} + 5 y_{d_g}^{-2} \right)}
         {\tfrac{T^4}{\Gamma_{s}} + \sum_g \left(y_{u_g}^{-2} + y_{d_g}^{-2}\right)} \,
    \frac{\dot{\theta}}{T} \, \Big|_{T_\mathrm{ew}} ,
    \label{eq:fin_B_val}
\end{equation}
where $g_{\star S}$ is the effective number of SM entropy degrees of freedom, 
$y_{u_g}$ ($y_{d_g}$) are the up- (down-)type Yukawa couplings for fermion generation $g$ that enter via the chirality--flipping process, and $\Gamma_s$ is the average strong sphaleron rate. All temperature-dependent quantities ($T$, $\Gamma_s$, $g_{\star S}, \dot\theta$) are evaluated at $T_\mathrm{ew}\sim 130\,{\rm GeV}$, the temperature at the onset of electroweak sphaleron freeze-out.  
Using the SM value for the several parameters, we find the observed baryon asymmetry \cite{Planck:2018vyg} is achieved for $\dot{\theta} (T_\mathrm{ew}) \simeq 4$~keV.

$ \langle\tilde{G}G \rangle_T$ also enters the axion equation of motion
\begin{align}\label{eqn:axion_EOM}
    \ddot\theta + 3H \dot\theta  = \frac{\alpha_c}{8\pi f_a^2} \langle \tilde{G} G \rangle_T  -\frac{1}{f_a^2}V'(\theta)  \, ,
\end{align}
where we have normalized the axion anomaly coupling as $\frac{\alpha_c}{8\pi} \, \frac{a}{f_a} \, G \tilde{G}$ with $f_a$ the PQ-breaking scale. At temperatures well-above confinement, the gauge field topological susceptibility is negligible and $V(\theta)$ vanishes. 
Putting (\ref{eq:sph_rate}) back in the equation of motion of Eq.~\eqref{eqn:axion_EOM}, we find a friction term for the axion parametrized by
\begin{equation}
    \Upsilon \;\equiv\; -\frac{\dot{n}_{\mathrm{CS}}}{\dot{\theta}} \, .
\end{equation}
The QCD sphaleron rate is studied in Refs.~\cite{Bonanno:2023thi,Bonanno:2023ljc}; throughout this \textit{Letter}, we use the lattice fit results of Ref.~\cite{Bonanno:2023thi} to calculate the friction for any confining force. Per the discussion in our companion paper \cite{Asadi:2026zbc}, this is a conservative modeling choice.

Solving the Boltzmann equations for the fermion asymmetries, we can calculate $\dot{n}_{\mathrm{CS}}$ in the electroweak-symmetric/broken phase to find \cite{Berghaus:2020ekh} 
\begin{eqnarray}
    \label{eq:ch_flip_rates1}
    \Upsilon 
    &=& \frac{\Gamma_{s}}{2 T}
        \left(1 + \frac{\Gamma_s}{T^4}\sum_f \, \left\{\begin{array}{cc}y_{f}^{-2}& T > T_{\rm ew}\\\frac{T^2}{\alpha_c m_f^2}&T < T_{\rm ew}\end{array}\right. \, \right)^{-1}.
\end{eqnarray} 
The factor of one corresponds to the friction exerted on the axion by a pure Yang--Mills theory, while fermions can only reduce the friction.

These equations show that the lightest fermions charged under the gauge group dominate the friction. 
This reflects a simple intuition. First, the fermion chiral asymmetry, if not washed out, will dynamically adjust to exactly cancel the effective chemical potential induced by the rolling axion. Second, only the fermion masses wash out the chiral asymmetry, so the species with the smallest masses---and thus the smallest flip rates---control the overall friction. 
This also illustrates that adding new heavy particles charged under QCD does not change the friction the QCD axion feels.

\begin{figure}
    \centering
    \includegraphics[width=\linewidth]{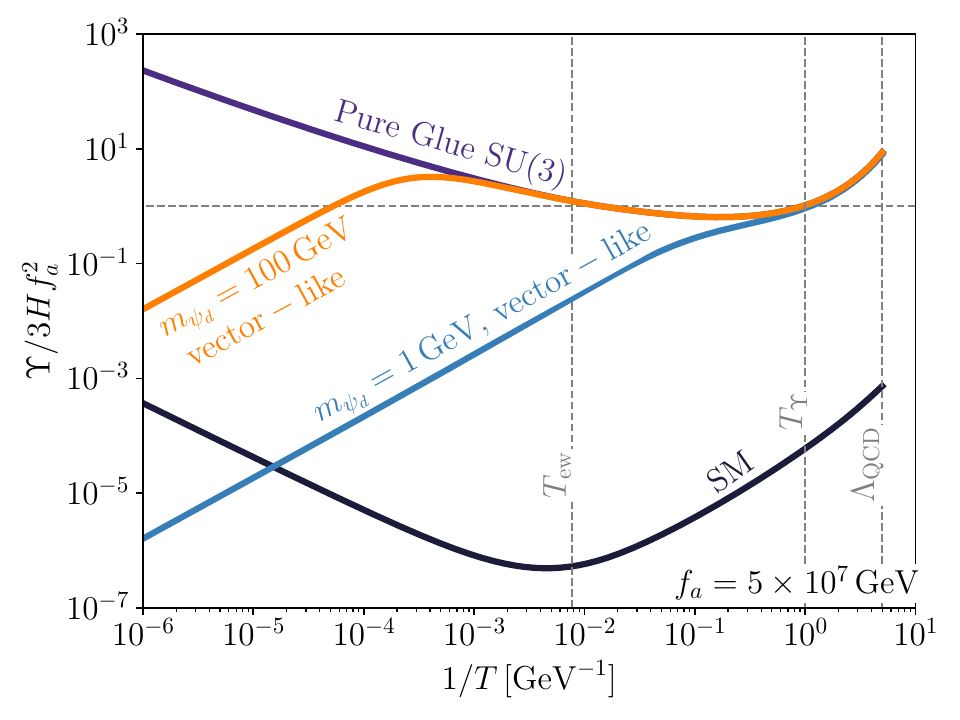}
    \caption{
    Friction felt by an axion and normalized by the Hubble rate as a function of the temperature in the SM (black curve), a pure Yang-Mills theory (\textcolor{DeepPurple}{purple} curve), and a confining SU(3) dark gauge group with a vector-like fermion of mass 100 GeV (\textcolor{HypotheticOrange}{orange} curve) or 1 GeV (\textcolor{HypotheticBlue}{blue} curve).
    All these theories are assumed to have the same confining scale as QCD. We fix the axion decay constant $f_a$ to the minimum value compatible with experimental bounds to maximize the friction; the overclosure problem is only exacerbated for higher values of $f_a$. The vertical lines illustrate relevant temperatures: the line labeled by $T_\Upsilon$ denotes the temperature at which the friction in a vector-like confining SU(3) dominates over Hubble below $T_{\rm ew}$, while the lines labeled $T_{\rm ew}$ and $\Lambda_{\rm QCD}$ indicate the electroweak and QCD phase transitions, respectively. The rapid increase of the sphaleron friction in the pure Yang–Mills (and heavy vectorlike fermion) case arises from the fast growth of the gauge coupling as the temperature approaches confinement, combined with the steep $\alpha$-scaling of the thermal sphaleron rate $\propto \alpha^5 T^4$~\cite{Arnold:1996dy}.}
    \label{fig:friction}
\end{figure}

In \cref{fig:friction} we show the friction felt by the QCD axion normalized by the Hubble rate as a function of the temperature, assuming a radiation-dominated cosmology. 
The temperature-scaling of the friction changes as we go through the electroweak phase transition, as indicated by Eq.~\eqref{eq:ch_flip_rates1}.
We find that in the SM, the friction felt by the axion is orders of magnitude smaller than the Hubble rate, i.e. QCD axion effectively feels no friction.

Lack of friction on QCD axions proves to be a major limitation for the original axiogenesis model. The axion continues rotating until the QCD epoch, leading to a relic abundance that overcloses the universe for values of $\dot{\theta}$ large enough to generate the observed baryon asymmetry today \cite{Co:2019wyp,Co:2019jts}.
Thus, without further ingredients, the original axiogenesis proposal can not explain both the baryon asymmetry of the universe and the observed DM abundance today.

Depending on whether chirality--flipping arises from a Yukawa coupling or from a (vector-like) mass, the temperature dependence of the friction changes.  
In \cref{fig:friction} we also show the friction predicted by a confining dark sector with a vector-like fermion with mass, $m_\psi=100$ GeV (\textcolor{HypotheticOrange}{orange} curve) or $m_\psi=1$ GeV (\textcolor{HypotheticBlue}{blue} curve). We find that the friction rapidly grows in the early universe and saturates the pure Yang-Mills limit (\textcolor{DeepPurple}{purple} curve) before the electroweak symmetry breaking in the case of a 100 GeV dark fermion. The saturation happens at later times for lighter dark fermions.

\vspace{0.1in}


\paragraph*{\textbf{Toy Model Axiverse ---}}
The overclosure problem can be avoided by introducing an efficient source of friction. As discussed above, a confining dark sector with heavy vector-like quarks can provide such friction and drain excess axion kinetic energy into a bath of dark gluons. However, this minimal extension introduces a new challenge: a generic confining force carries its own strong CP phase, uncorrelated with that of QCD. If a single axion couples to both gauge groups, the resulting combined potential is generically minimized away from the CP-even point, and no longer solves the strong CP problem.

More broadly, sourcing a large axion rotational velocity through explicit PQ breaking can introduce a tension with the high PQ quality required for the strong CP solution. Additional PQ symmetries offer a natural way to avoid this tension by allowing sizeable explicit breaking in directions orthogonal to the QCD axion.

Both these shortcomings are naturally resolved in an (open \cite{Petrossian-Byrne:2025mto}) string axiverse \cite{Witten:1984dg,Arvanitaki:2009fg} and with multiple axions.\footnote{Similar setups are studied in Refs.~\cite{Cheng:2001ys,Loladze:2025uvf}.}  
In a multi-axion framework, the QCD axion is typically a linear combination of many pseudoscalars, and only one of them must possess the high-quality PQ symmetry needed to solve the strong CP problem. 
Extra-dimensional axions further allow different modes to exhibit different levels of PQ quality; see Ref.~\cite{Reece:2025thc} for a recent review. Consequently it is plausible that only a subset of axions experience large PQ breaking (and are therefore spun up), while others retain the required quality to solve the strong CP problem.

We consider a simplified ``axiverse'' with two axions:
\begin{equation}
\begin{aligned}\label{eqn:two_axion_Lagrangian}
    \mathcal{L} & \supset \frac12 f_1^2\partial_\mu\theta_1\partial^\mu\theta_1 +  \frac12 f_2^2\partial_\mu\theta_2\partial^\mu\theta_2 \\
    &+ \frac{\alpha_c}{8\pi}(\theta_1 + C\theta_2)\tr \tilde G G+\frac{\alpha_d}{8\pi}\theta_2\tr \tilde G_d G_d\, \\
    &+ \bar \psi_d\,( {\rm i}\slashed{D}_{d} - m_{\psi_d}) \,\psi_d + \mathcal{L}_\mathrm{portal} \, ,
\end{aligned}
\end{equation}
where $G_d$ is the new dark confining gauge group field strength, dark quark $\psi_d$ is responsible for generating a suitable friction on $\theta_2$ axion, and $\mathcal{L}_\mathrm{portal}$ captures further portal interactions to the SM (see below). 
This represents the most general infrared Lagrangian of two axion fields coupled to two non-Abelian gauge sectors.

The Lagrangian in Eq.~\eqref{eqn:two_axion_Lagrangian} can arise in several ways, including the KSVZ-like construction of Ref.~\cite{Cyncynates:2023esj}. A natural alternative is that both axions originate from an extra-dimensional or string-theoretic axiverse. In such setups, the QCD axion is typically a linear combination of many underlying fields. 
The PQ breaking for each axion depends exponentially on a distinct bulk mass in the extra dimension \cite{Kallosh:1995hi,Petrossian-Byrne:2025mto,Reece:2025thc}. Thus, while a single high-quality field ($\theta_1$) suffices to solve strong CP, additional fields (e.g., $\theta_2$) can have a much larger PQ-breaking, which triggers their spinning. 
We leave a detailed study of the origin of the axion rotations at $T \gg T_\mathrm{ew}$ to our companion paper \cite{Asadi:2026zbc}; here we follow their dynamics below the electroweak scale and show that the theory in \cref{eqn:two_axion_Lagrangian} could solve the overclosure problem.

This structure leads to initial conditions with $\dot\theta_1 (T_\mathrm{ew})\simeq 0$ while $|\dot\theta_2 (T_\mathrm{ew})|\gg 0$ via the kinetic misalignment mechanism~\cite{Co:2019jts}. The large radial excursion of the axion parent field invoked in Refs.~\cite{Co:2019wyp,Co:2019jts} can be replaced by a large \textit{axion} excursion in a \textit{monodromy potential} to yield a similar initial condition at the electroweak scale;
see our companion work~\cite{Asadi:2026zbc} for further details on this UV completion.

From this initial condition, the universe cools to the electroweak phase transition. 
Even though QCD sphalerons are inefficient at dissipating the energy density in $\theta_2$, they do transfer some energy to $\theta_1$, though $\dot\theta_1$ remains much smaller than $\dot\theta_2$ during the electroweak phase transition. 
Thus, at the temperature of electroweak sphaleron freeze-out ($T_\mathrm{ew}$), we impose
\begin{align}\label{eqn:Initial_Condition}
    \dot\theta_1  \approx 0\,,\hspace{1cm}C\dot\theta_2 \approx 4{\,\rm keV}\,,
\end{align}
to match the observed baryon abundance~\cite{Co:2019wyp}. 

\begin{figure}
    \centering
    \resizebox{\columnwidth}{!}{
    \includegraphics[width=0.5\linewidth]{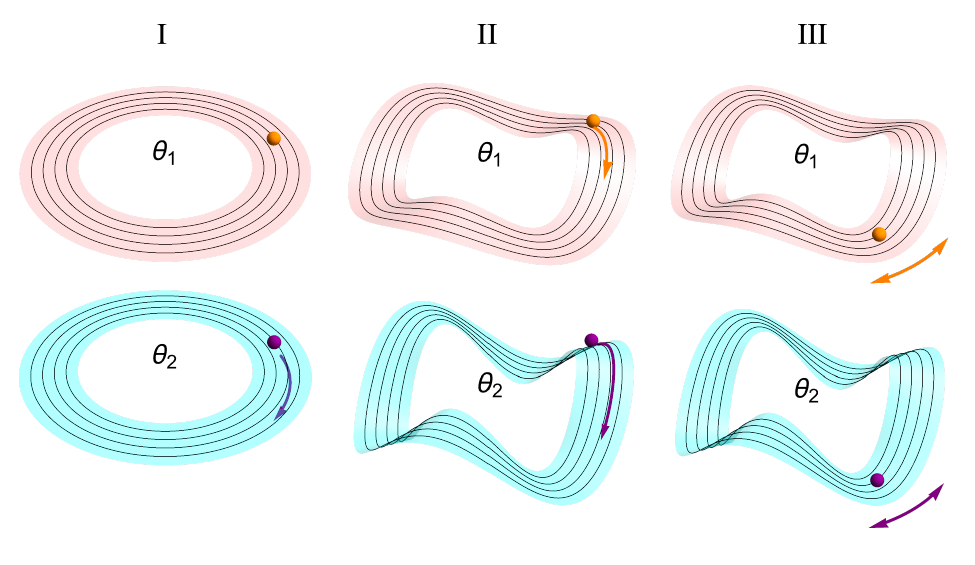}
    }
    \caption{Schematic presentation of the evolution of the two axions in their potentials below $T_\mathrm{ew}$ (from left to right). The friction to the dark sector takes away enough energy out of the system to evade the overclosure - see the text for details.
    }
    \label{fig:epochs}
\end{figure}

We solve the coupled equations of motion of the axions to find their evolution below $T_\mathrm{ew}$, schematically shown in \cref{fig:epochs}.\footnote{ In this \textit{Letter}, we neglect spatial fluctuations, which may be generated through parametric resonance~\cite{Fonseca:2019ypl} and may modify the resulting relic abundances. See Ref.~\cite{Asadi:2026zbc} for further details. } 
Starting from the initial condition in Eq.~\eqref{eqn:Initial_Condition} at $T_{\mathrm{ew}}$ (panel I in \cref{fig:epochs}),
since the dark quarks are vector-like and have a large mass, the dark friction grows as the universe cools (see \cref{fig:friction}) and eventually slows the rotation of the $\theta_{2}$ axion when the friction $\Upsilon_d/f_2^2$ becomes comparable to $3H$ at $T \sim T_\Upsilon$. As the temperature approaches the two confinement scales, the gauge groups generate potentials for both axions, coupling them through QCD susceptibility (panel II); at this point both axions have enough kinetic energy to rotate.  
Eventually, enough energy is dissipated via the dark friction that the axions are trapped around their potential minimum and start oscillating instead of rotating (panel III).

The axions' kinetic energy is transferred to dark gluons, which later hadronize into glueballs once the dark sector confines. Without additional dynamics, these relic glueballs would overclose the universe. Glueballs may then decay back to the SM via loops of heavy vector-like fermions $\xi_d$ that are bifundamentals of SM SU(2)$_L$ and the dark confining group.
The heavy dark quarks $\xi_d$ can form dark baryons that constitute only a tiny fraction of the dark matter density \cite{Mitridate:2017oky} and produce signals at colliders \cite{Asadi:2025vfr},\footnote{If $\psi_d$ does not transform as fundamentals of the new confining force but $\xi_d$ does, the abundance of dark baryons may further dilute through dark matter squeeze-out \cite{Asadi:2021yml,Asadi:2021pwo,Asadi:2022vkc} if the dark confinement phase transition is first order; while acknowledging this possibility, we do not pursue it further here.} while their direct-detection signatures can be completely suppressed \cite{Asadi:2024bbq,Asadi:2024tpu}.
Moreover, given the existing LHC bounds on the $\xi_d$ mass \cite{Asadi:2025vfr}, the dark confinement scale, and therefore the mass of $\theta_2$, is bounded from below to avoid the stringent BBN limits \cite{Kawasaki:2020qxm} on late glueball decays \cite{Juknevich:2009ji,Juknevich:2009gg,Forestell:2016qhc,Soni:2016gzf,Forestell:2017wov,McKeen:2024trt}. 

\begin{figure}
    \centering
    \includegraphics[width = \columnwidth]{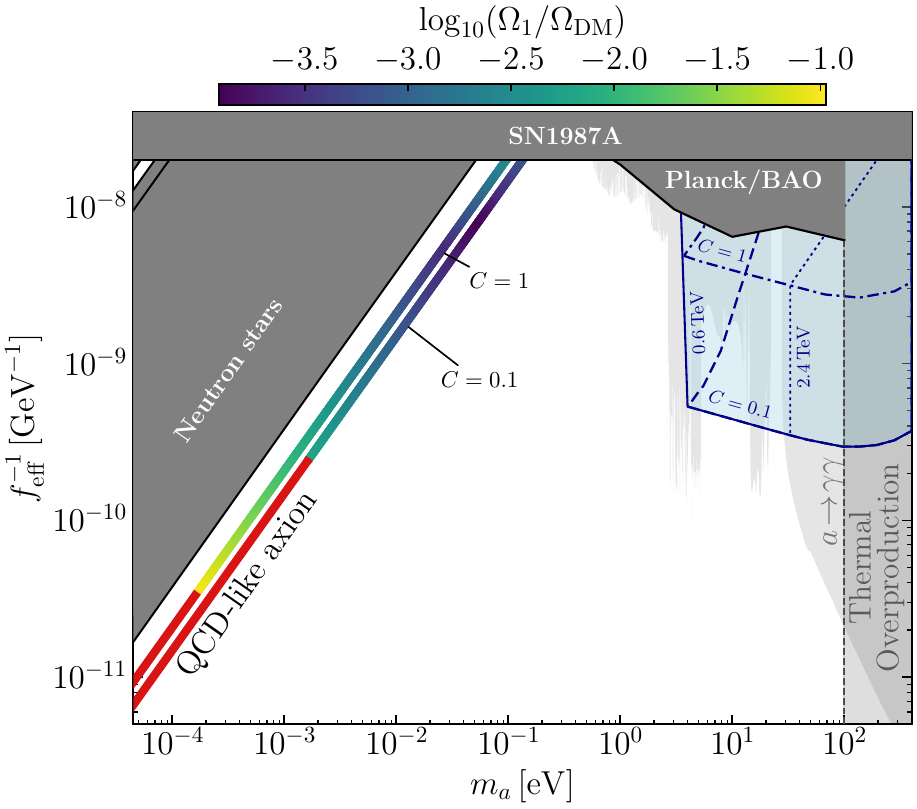}
    \caption{Parameter space (mass $m_i$ vs. QCD coupling scale $f_{\rm eff}$) for $\theta_1$ (colored lines) and $\theta_2$ (\textcolor{HypotheticBlue}{blue} regions) consistent with avoiding overclosure.
    The left boundary of the \textcolor{HypotheticBlue}{blue} region is set by requiring $\Lambda_d>\Lambda_{\rm QCD}$ and sufficiently rapid glueball decay for the indicated bifundamental fermion masses (600 GeV, 2.4 TeV in the figure) due to glueball lifetime constraints. 600 GeV is roughly the present LHC bound. Note that the bifundamental fermion is separate from the lighter dark-charged fermion responsible for suppressing dark friction at high temperatures. The different linestyles correspond to (solid) the set of allowed regions for $C>0.1$, (dashed) the allowed region with $C = 0.1$ and over all allowed bifundamental quark masses, (dot-dashed) the same for $C = 1$, and (dotted) the allowed region with a bifundamental quark mass larger than 2.4 TeV. The lower boundary of these regions is determined by the requirement that $\theta_2$ not overclose the universe, while the kink around 100 eV arises from the requirement that sphaleron friction is subdominant to Hubble friction above $T_{\rm ew}$. The bounds in dark gray rely on the axion-gluon coupling~\cite{Hook:2017psm,Gomez-Banon:2024oux,Kumamoto:2024wjd,Blum:2014vsa,Springmann:2024ret,Balkin:2022qer,Caloni:2022uya} (the ``Thermal Overproduction'' bound depends only on the gluon coupling). The bounds in light gray assume a coupling of $\theta_2$ to photons given by $\frac{\alpha_{\rm EM}}{2\pi}\theta_2 F\tilde F$~\cite{Yin:2024lla,Regis:2020fhw,Todarello:2023hdk,Saha:2025any,Grin:2006aw,Janish:2023kvi,Wadekar:2021qae,Blout:2000uc,Carenza:2023qxh,Nakayama:2022jza,Todarello:2024qci,Wang:2023imi,Cadamuro:2011fd,Capozzi:2023xie,Liu:2023nct} and that it constitutes 100\% of the dark matter.  The colored QCD axion lines show the $\theta_1$ relic abundance from ordinary misalignment and transfer from $\theta_2$ via QCD sphaleron friction, assuming an early period of $\theta_2$ kinetic domination and $f_1=f_2$. Red segments indicate $\theta_2$ kinetic domination at $T_{\rm EW}$.}
    \label{fig:param_space}
\end{figure}

The region of parameter space consistent with both the observed baryon asymmetry and the dark matter relic abundance is shown in \cref{fig:param_space}. The blue region indicates where the total axion energy density is at or below the observed dark matter abundance, and there exist parameters where every point in this region can account for the entirety of the dark matter. To determine this region, we numerically evolve the axion equations of motion from the boundary condition in \cref{eqn:Initial_Condition}, using the WKB approximation where applicable to compute the present-day abundances of the axion mass eigenstates.
The QCD axion abundance also depends on the cosmology preceding the electroweak phase transition. Starting from $\dot\theta_1=0$ at $\theta_2$ kinetic–radiation equality, QCD sphaleron friction generates a rotation of $\theta_1$. We estimate the resulting relic abundance analytically, including ordinary misalignment averaged over a uniform initial angle:
\begin{equation}
\begin{aligned}
    \frac{\langle\Omega_1\rangle}{\Omega_{\rm DM}}&\simeq \max\left\{10^{-3}|C|\left(\frac{f_1}{10^8\,{\rm GeV}}\right)^{-2}\frac{f_1}{f_2},\right.\\
    &\left.\hspace{2.5cm}10^{-4}\left(\frac{f_1}{10^8\,{\rm GeV}}\right)^{7/6}\right\}.
\end{aligned}
\label{eq:abundance_1}
\end{equation}
These estimates are illustrated by the color gradient along the QCD axion lines in \cref{fig:param_space} for $f_1=f_2$. 
At small $f_1$, the abundance is dominated by friction-induced kinetic misalignment (the first term), while at large $f_1$, ordinary misalignment dominates (the second term).

This framework also admits direct detection signals in axion dark matter experiments 
and astrophysical searches. Given the mass range accessible to $\theta_2$, the most 
promising search strategy is its two-photon decay in telescope observations~\cite{Yin:2024lla,Regis:2020fhw,Todarello:2023hdk,Saha:2025any,Grin:2006aw,Janish:2023kvi,Wadekar:2021qae,Blout:2000uc,Carenza:2023qxh,Nakayama:2022jza,Todarello:2024qci,Wang:2023imi,Cadamuro:2011fd,Capozzi:2023xie,Liu:2023nct}.

The $\theta_1$ axion itself may be probed in direct detection experiments \cite{Baryakhtar:2018doz,IAXO:2019mpb,BREAD:2021tpx,Batllori:2023gwy}, particularly 
near the upper end of its mass range around $0.1\,\mathrm{eV}$ where its abundance arises from kinetic misalignment. 
Rapid glueball decay requires an upper bound on the mass of bifundamental fermions $\xi_d$, placing them within reach of HL-LHC and future colliders. 

\vspace{0.2in}

\textbf{Note Added.} While completing this work, Ref.~\cite{Co:2025lcs} appeared on arXiv that provided a different resolution for one of the shortcomings of the original axiogenesis framework (the overclosure problem) in a different setup.

\section*{ACKNOWLEDGMENTS}
We thank Masha Baryakhtar, Arushi Bodas, Rikab Gambhir, Samuel Homiller, Anson Hook, Junwu Huang, Gordan Krnjaic, Guy Moore, Jakob Moritz, Matt Reece, Nick Rodd, and Jure Zupan for helpful discussions. 
P.A{.} thanks Brian Batell, Jae Hyeok Chang, Gongjun Choi, Arnab Dasgupta, Ayres Freitas, Tony Gherghetta, Ameen Ismail, Seth Koren, Zhen Liu, Robert McGehee, Zahra Tabrizi, Yuhsin Tsai, Lian-Tao Wang, and Ho-Ung Yee, for their illuminating questions during the presentations of a preliminary version of this \textit{Letter}. 
The research of P.A{.} and S.G{.} is supported in part by the U.S{.} Department of Energy grant number DE-SC0010107. D.C{.} acknowledges the receipt of the grant from the Abdus Salam International Centre for Theoretical Physics (ICTP), Trieste, Italy. D.C{.} also acknowledges support from the Department of
Physics and College of Arts and Science at the University
of Washington and the U.S{.} Department of Energy
Office of Science under Award Number DE-SC0024375.
This work was initiated at the Aspen Center for Physics, which is supported by National Science Foundation grant PHY-2210452. This research was supported in part by grant NSF PHY-2309135 to the Kavli Institute for Theoretical Physics (KITP). Experimental limits presented in this paper were in part compiled using the GitHub repository associated with Ref.~\cite{AxionLimits}.

Generative AI tools, including ChatGPT versions 4 through 6 and Claude Fable, were used for language editing, code assistance, and to confirm mathematical results. The authors take full responsibility for the content of the manuscript.

\clearpage

\bibliographystyle{apsrev4-1}
\bibliography{bibliography}

\end{document}